%% file: pre_manuscript.tex
\documentclass[aps,twocolumn,pre]{revtex4-1} 
\usepackage[utf8]{inputenc}
\usepackage[english]{babel}
\usepackage{amsmath}
\usepackage{amsfonts}
\usepackage{amssymb}
\usepackage{graphicx}
\usepackage{array}
\usepackage{bm}
\usepackage{xcolor}
\usepackage[colorlinks=true,linkcolor=blue,citecolor=blue,filecolor=cyan,urlcolor=blue,breaklinks=true]{hyperref}

\newcommand{\fext}{F_\text{ext}}
\newcommand{\ps}{x, v}
\newcommand{\psVec}{\bm{\xi}}
\newcommand{\psInt}{\int dv \int dx \,}
\newcommand{\uu}[1]{ \bm{#1}}
\newcommand{\mean}[1]{\left\langle #1 \right\rangle}
\newcommand{\kT}{T}
\newcommand{\TT}{\mathcal{T}} 

\begin{document}

\title{Large Deviation Function for a Driven Underdamped Particle in a Periodic Potential}
\author{Lukas P. Fischer}
\author{Patrick Pietzonka}
\author{Udo Seifert}
\affiliation{II. Institut f\"ur Theoretische Physik, Universit\"at Stuttgart, 70550 Stuttgart, Germany}


\begin{abstract}
Employing large deviation theory, we explore current 
fluctuations of underdamped Brownian motion for the
paradigmatic example of a single particle in a one dimensional
periodic potential. Two different approaches to the large
deviation function of the particle current are presented.
First, we derive an explicit expression for the large deviation
functional of the empirical phase space density, which replaces
the level 2.5 functional used for overdamped dynamics. Using
this approach, we obtain several bounds on the large deviation
function of the particle current. We compare these to bounds for
overdamped dynamics that have recently been derived motivated by the 
thermodynamic uncertainty relation. Second, we provide a
method to calculate the large deviation function via the
cumulant generating function. We use this method to assess the
tightness of the bounds in a numerical case study for a cosine
potential.

\end{abstract}

\maketitle

\section{Introduction}
The driven Brownian particle in a periodic potential is a paradigmatic model
that can be mapped to a wide class of systems,
notably Josephson junctions \cite{buet83}, molecular motors
\cite{juel97, haya15, shin16} and colloidal systems \cite{fauc95,blic07,ma17}. In these contexts, the model 
has repeatedly been used to explore the interplay between, on the one hand,
the driving force and the shape of the potential and, on the other hand, the
particle current and its fluctuations. One striking result is the enhancement
of the effective diffusion coefficient by several orders of magnitude in a
specific range of driving forces \cite{reim01}.

On long time scales, the fluctuating particle current is directly connected to
the entropy production and its fluctuations, which plays a central role
in stochastic thermodynamics \cite{seif12}. The arguably best framework for
studying fluctuations on large time scales is provided by the theory of large
deviations, which defines the so-called large deviation function that
characterises the exponential decay in the probability of atypical
fluctuations \cite{touc09, laza15}. In this description, the fluctuation theorem for
entropy production is expressed as a symmetry of the corresponding large
deviation function \cite{lebo99}. For the important special case of overdamped
Brownian dynamics in periodic potentials, this and other characteristic
properties of the large deviation function, such as the emergence of a kink at
zero entropy production, have been comprehensively discussed
\cite{mehl08,doro11, budi11a,spec12,tsob16}.

Underdamped dynamics, where damping does not suppress inertial effects, is
conceptually important for stochastic thermodynamics, as it is closer to a
microscopic Hamiltonian description and shows different effects
\cite{blan98,cela12,ge14}. So far, however, much less is known about the properties of
large deviation functions for currents in the underdamped regime compared to the 
overdamped regime. Only very recently, large deviation theory has been
considered for underdamped Langevin dynamics in a rather general,
mathematical setting \cite{jack17}. In this paper, we derive an explicit
expression for the level 2 large deviation function for underdamped
dynamics in a periodic potential. This function describes the probability to
observe a specific empirical phase space density and replaces the expression
for the level 2.5 large deviation function for the joint probability of
empirical densities and currents typically encountered in the  overdamped regime
\cite{maes08b, bara15d, hopp16}. Moreover, we derive the calculation of the large deviation
function for the particle current via Legendre transformation of the cumulant
generating function.

The development of large deviation theory for underdamped Brownian motion
appears particularly relevant in the light of the thermodynamic uncertainty
relation, which expresses a universal trade-off between desired small
fluctuations of a current and the rate of entropy production
\cite{bara15}. For Markov processes on discrete state-spaces and overdamped
Brownian motion, this relation can be proven as a consequence of thermodynamic
bounds on the large deviation function
\cite{piet15,ging16,pole16,ging16a,garr17,proe17,piet17,horo17,maes17,dech17,nard17a, hyeo17},
a more direct proof using martingale theory might be possible as a special
case has shown \cite{pigo17}. However, naively applying these methods to
underdamped dynamics leads to a formally divergent contribution to the
entropy production from the reversible phase space currents. A recent study
involving piecewise ballistic processes even suggests that the uncertainty
relation might not hold for underdamped dynamics in the presence of magnetic
fields \cite{bran18}. Motivated by these results, we explore the bounds on
the large deviation function of the particle current that follow from various ansatzes
for the empirical density in our expression for the level 2 large deviation
function. 

The paper is organised as follows. For the setup described in
Sec.~\ref{sec:setup}, we derive the level 2 large deviation theory in
Sec.~\ref{sec:level2} and the cumulant generating function in
Sec.~\ref{sec:scgf}. The latter method is used for a numerical case study for
a cosine potential in Sec.~\ref{sec:numerics}. In Sec.~\ref{sec:bounds},
several bounds on the large deviation function are derived. A particular bound
and its connection to the thermodynamic uncertainty relation is discussed in
Sec.~\ref{sec:tdur}. We conclude in Sec.~\ref{sec:conclusion}.


\section{Setup}
\label{sec:setup}
We consider a Brownian particle with mass $m$ on a ring with radius $R$ in a
$2\pi R$-periodic potential $V(x)$, where $x$ is the position
of the particle on the ring. The particle with friction coefficient $\gamma$
is immersed in a solvent with temperature $T$. An additional external force
$\fext$ drives the system out of equilibrium. The underdamped dynamics of the
system is described by the Langevin equation
\begin{align}
	\dot{x}(t) &= v(t) \nonumber \\
	\label{eq:langevinUnderdamped} m\dot{v}(t) &= - \gamma v(t) + \fext - V'(x(t)) +  \eta(t)
\end{align}
with periodic boundary conditions for $x\in[0, 2\pi R)$.
The stochastic force $\eta(t)$ is Gaussian white noise with zero mean and correlations $\langle
\eta(t) \eta(t')\rangle = 2T\gamma \delta(t-t')$, where $\langle \cdot\rangle$ denotes
an average over all realisations. Throughout the paper, we set
Boltzmann's constant $k_\mathrm{B}=1$. The fluctuating internal energy of the system is
given by
\begin{equation}
  \label{eq:Eint}
  E(\ps)\equiv V(x)+\frac{1}{2}m v^2.
\end{equation}

Our aim is to characterise the statistical properties of the fluctuating
current 
\begin{equation}
  \label{eq:intcurr}
  J_\TT\equiv\frac{1}{\TT} \int_0^\TT dt\, v(t),
\end{equation}
which measures the accumulated distance travelled by the particle along an individual
trajectory divided by the length $\TT$ of the latter.

The evolution of the corresponding distribution $p(\ps, t)$ in phase space is
given by the underdamped Fokker-Planck equation, also known as Kramers
equation,
\begin{align}
	\label{eq:FokkerPlanck}  	\partial_t p(\ps, t) = -\partial_x j_x(\ps, t) -\partial_v j_v(\ps, t) \equiv \mathcal{L}p(\ps, t)
\end{align}
with the microscopic currents 
\begin{align}
	\label{eq:jx}j_x(\ps, t) = & v p(\ps, t) \\
	\label{eq:jv}j_v(\ps, t)             =& - \frac{1}{m} \left( v \gamma  - \fext + V'(x)\right) p(\ps, t) \nonumber \\
															     &- \frac{\kT \gamma}{m^2} \partial_v p(\ps, t) ,
\end{align}
and where we identify the Fokker-Planck operator $\mathcal{L}$.

In the following, we focus exclusively on the steady state,
where the time derivative in the Fokker-Planck equation
\eqref{eq:FokkerPlanck} vanishes, thus defining the stationary distribution
$p^\text{s}(\ps)$ and corresponding phase-space currents $j_x^\text{s}(\ps)$
and $j_v^\text{s}(\ps)$. The mean particle current on the ring follows
readily as
\begin{equation}
  \label{eq:js}
  J^\text{s}\equiv\mean{J_\TT}=\mean{v(t)}=\psInt  j^\textup{s}_x(\ps),
\end{equation}
where phase-space integrals have boundaries $v\in(-\infty,+\infty)$ and
$x\in[0,2\pi R)$ throughout the paper.

The non-equilibrium character of the driven system is quantified by its rate
of entropy production. Along an individual phase-space trajectory $\psVec(t) =
(x(t),v(t))$ of length $\TT$, the change in total entropy is defined as \cite{seif12}
\begin{equation}
	\label{eq:entropy} \varDelta s_\text{tot}[\psVec(t)] \equiv \ln \frac{\mathcal{P}[\psVec(t)]}{\mathcal{P}[\tilde{\psVec}(t)]} ,
\end{equation}
which compares the probabilities of a specific trajectory $\psVec(t)$ and its
time-reversed counterpart $\tilde{\psVec}(t)\equiv (x(\TT-t),-v(\TT-t))$. 
In the steady state, the rate of entropy production becomes \cite{spin12}
\begin{equation}
	\sigma\equiv \label{eq:meanEntropyProduction1} \langle \dot{s}_\text{tot}(t) \rangle =\frac{m^2}{\gamma \kT} \psInt \frac{j_v^\text{ir}(x, v, t)^2}{p(x, v, t)} \geq 0,
\end{equation}
with the stationary irreversible current 
\begin{equation}
	\label{eq:jvIR} j_v^\text{ir}(\ps) \equiv -\frac{\gamma v}{m} p^\text{s}(\ps) - \frac{\kT\gamma}{m^2} \partial_v p^\text{s}(\ps) .
\end{equation}
Through partial integrations, the rate of entropy production can be written as
\begin{equation}
  \label{eq:meanEntropyProduction2}
  \sigma=\fext J^\text{s}/T,
\end{equation}
which expresses that, due to conservation of energy, the average work
performed by the external force matches the average dissipated heat. Indeed,
even along individual trajectories, both work and dissipated heat are equal
to the total entropy production up to finite contributions from internal
energy and entropy that do not scale with the time $\TT$.

\section{Level 2 large deviation theory}
\label{sec:level2}

The probability to observe some current $J_\TT=J$ for a trajectory of length $\TT$ follows a so called large deviation principle \cite{touc09}
\begin{equation}
	\label{eq:ldPrinciple} P(J; \TT) \approx e^{-\TT I(J)}
\end{equation}
with the large deviation function (LDF) $I(J)$. This function captures the
exponential rate of decay of the probability to observe a specific current $J$
for large times $\TT$ with $I(J) \geq 0$ and equality only for $J = J^\text{s}$. Large deviation functions for such scalar, time-additive observables are called \emph{level 1} LDF.

For the so-called \emph{level 2} of large deviation theory, the object of interest is the fluctuating empirical distribution in the state space. In our instance of a two-dimensional phase space, we consider the empirical phase space density, defined as
\begin{equation}
	\label{eq:empiricalDensity} \rho (\ps) \equiv \frac{1}{\TT} \int_0^\TT dt\, \delta(x-x(t))\delta(v-v(t)) 
\end{equation}
along an individual trajectory $(x(t),v(t))$. The exponential decay of the
probability to observe a specific realisation of the fluctuating scalar field $\rho$
is described by the functional \emph{level 2} LDF $I[\rho]$.
Beyond that, the \emph{level 3} LDF considers the decay of the probability for
realisations in the trajectory space.

On an intermediate level, dubbed \emph{level 2.5}, the LDF $I[\rho, \bm\mu ]$ measures the decay of 
joint probability for a specific empirical density \eqref{eq:empiricalDensity} and an empirical current, which we define for the two-dimensional phase space as
\begin{equation}
	\label{eq:empiricalCurrent} \bm\mu (\ps) \equiv  \frac{1}{\TT} \int_0^\TT dt\,
	\delta(x-x(t))\delta(v-v(t)) \left(	\begin{matrix}
		\dot{x}(t) \\
		\dot{v}(t)
	\end{matrix} \right) 
\end{equation}
with components $\mu_x(\ps)$ and $\mu_v(\ps)$.  For overdamped Brownian motion and Markovian networks, this level of large deviation theory has proven as the appropriate description in non-equilibrium systems. In contrast to level 2, the level 2.5 LDF contains statistics on the currents in the system and thus allows one to  infer the level 1 LDF for all types of time-additive observables \cite{maes08b, bara15d, hopp16, ging16a}, including current-like variables such as the entropy production.

In order to derive
an expression for the LDF for underdamped dynamics, we first consider the auxiliary dynamics
\begin{align}
	\label{eq:app2AuxDynamics} \dot{x}(t) &= v(t) + \sqrt{ \epsilon } \tilde{\eta}(t) \nonumber \\
	m\dot{v}(t) &= -\gamma v(t) + \fext - V'(x(t)) + \eta(t),
\end{align}
with an additional Gaussian white noise with zero mean,
$\mean{\tilde\eta(t)\tilde\eta(t')}=2\delta(t-t')$, and that is uncorrelated
to $\eta(t)$. This dynamics corresponds effectively to an overdamped diffusion in the
two-dimensional $x$-$v$-space with a diagonal diffusion matrix and reproduces the
original Langevin equation \eqref{eq:langevinUnderdamped} in the limit
$\epsilon \rightarrow 0$. 

With both coordinates being diffusive, the level 1 LDF for the current $J$ can be derived within the framework of level 2.5 large deviations through the contraction principle
\begin{equation}
	\label{eq:level25LDF_contraction} I(J) = \mspace{-50mu} \min_{  \mspace{+60mu}   \rho, \bm{\mu} \; |\;  J = J[\bm\mu ]} I[\rho, \bm{\mu}].
\end{equation}
Here, we can associate a macroscopic current for every choice of the empirical current through the functional
\begin{equation}
	J[\bm\mu] \equiv \psInt \mu_x(x, v). 
\end{equation}
The contraction condition $J = J[\bm\mu]$ in
Eq.~\eqref{eq:level25LDF_contraction} thus ensures that the macroscopic
current $J$ matches the empirical densities. The level 2.5 LDF for the
auxiliary dynamics \eqref{eq:app2AuxDynamics} is obtained from the
general, explicit expression for multidimensional overdamped diffusion
as
\begin{align}
	\label{eq:app2LDF25}	
	I[\rho, \bm{\mu}] = \frac{1}{4} \psInt &\frac{1}{\rho \epsilon}\left( \mu_x - j_x^\rho \right)^2  \nonumber \\
	                                   & + \frac{m^2}{\gamma \kT \rho}\left( \mu_v - j_v^\rho\right)^2,
\end{align}
if $\nabla\bm{\mu} = 0$, and is otherwise assigned the value $\infty$. Note
that we omit the arguments of all phase-space functions in the integral for better readability.
The currents
\begin{equation}
  j^\rho_x(\ps) \equiv v \rho(\ps) - \epsilon \partial_x \rho(\ps) 
\end{equation}
and
\begin{equation}
  j^\rho_v(\ps) \equiv - \frac{1}{m} \left( v \gamma  - \fext +
    V'(x)+\frac{T\gamma}{m}\partial_v\right) \rho(\ps)
\label{eq:jrhov}
\end{equation}
entering \eqref{eq:app2LDF25} are the Fokker-Planck currents corresponding to
an empirical density $\rho(\ps)$ and the auxiliary dynamics \eqref{eq:app2AuxDynamics}.

Simplification of \eqref{eq:app2LDF25} and expansion in $\epsilon$ yields
\begin{align}
	\label{eq:app2LDF25-2}I[\rho, \bm{\mu}] = \frac{1}{4} \psInt \frac{1}{\epsilon \rho} \left( \mu_x -  v\rho \right)^2  + \mathcal{O}( 1 ) .
\end{align}
Returning to underdamped dynamics \eqref{eq:langevinUnderdamped} by considering
the limit $\epsilon \rightarrow 0$, the integral diverges unless the empirical
density and current satisfy the constraint
\begin{equation}
	\label{eq:underdampedMuXFix} \mu_x(\ps) = v\rho(\ps).
\end{equation}
Together with the constraint of the current being divergence-free,
\begin{equation}
	\label{eq:empiricalCurrentDiv} \nabla \bm{\mu}(\ps)=\partial_x \mu_x(\ps) + \partial_v \mu_v(\ps) = 0 ,
\end{equation}
we then find as a constraint on $\mu_v(\ps)$
\begin{align}
	\label{eq:underdampedEmpiricalCurrent}\mu_v(\ps)= -\int_{-\infty}^vdv'\, v' \partial_x \rho(x, v') \equiv \mu_v^\rho(\ps)  ,
\end{align}
where we have used that the empirical current must vanish for $v \rightarrow
\pm \infty$ for $I[\rho,\bm{\mu}]$ to be finite. As a consequence, the
empirical current $\bm{\mu}(\ps)$ is completely fixed by the choice of the
empirical density $\rho(\ps)$. The contraction \eqref{eq:level25LDF_contraction} then reduces to a minimisation 
\begin{equation}
	\label{eq:underdampedLevel20Contraction} I(J) =\mspace{-36mu} \min_{ \mspace{42mu} \rho \; |\;  J =\langle v \rangle_\rho} I[\rho]
\end{equation}
over only the empirical density $\rho (\ps)$ with the constraint that the particle current equals
\begin{equation}
  \label{eq:level2current}
  J=\langle v \rangle_\rho\equiv \psInt dv\, v\rho(\ps) .
\end{equation}
Moreover, we have the implicit constraints that $\rho(\ps )$ must be normalized and that the microscopic current $\mu^\rho_v (x, v)$ must  vanish for $v\rightarrow \pm \infty$. The latter constraint implies through Eq. \eqref{eq:underdampedEmpiricalCurrent} that the particle current is constant along the ring, such that the $x$-integration in Eq. \eqref{eq:level2current} leads to a prefactor $2\pi R$ .
As our main result, the large deviation functional
\begin{equation}
	\label{eq:underdampedLevel20LDF} I[\rho] = \frac{m^2}{4 \gamma \kT} \psInt \left( \mu_v^\rho - j_v^\rho\right)^2/\rho  ,
\end{equation}
with $j_v^\rho(\ps)$ and $\mu_v^\rho(\ps)$ as defined in Eqs. \eqref{eq:jrhov}
and \eqref{eq:underdampedEmpiricalCurrent} then follows directly from Eq.~\eqref{eq:app2LDF25}. 

In conclusion, we observe that for underdamped dynamics, the level 2.5 of large
deviation theory degenerates to level 2 involving only an empirical phase
space density $\rho(\ps)$. Nevertheless, one can contract to the level 1 LDF for the current, since the fluctuating current, Eq. \eqref{eq:intcurr}, is the average of a scalar field along the trajectory for underdamped dynamics.

\section{Tilted operator and cumulant generating function}
\label{sec:scgf}

As stated by the
G\"artner-Ellis theorem, the LDF for the empirical particle current $J_\TT$,
Eq.~\eqref{eq:intcurr}, is connected to the scaled cumulant generating
function \citep{touc09}
\begin{equation}
	\label{eq:defGenFunc} \alpha(\lambda) \equiv \lim_{\TT \rightarrow \infty} \frac{1}{\TT}\ln \langle e^{\lambda \TT J_\TT } \rangle
\end{equation}
with a real parameter $\lambda$ via the Legendre-Fenchel transformation
\begin{equation}
	\label{eq:gaertnerEllis} I(J) = \max_{\lambda}\left[ \lambda J - \alpha(\lambda) \right] .
\end{equation}
The derivatives of $\alpha(\lambda)$ at $\lambda=0$, ``generate'' the cumulants
of the distance travelled in a large time $\TT$. In particular, the average
particle current \eqref{eq:js} is $J^\text{s}=\alpha'(0)$ and its effective
diffusion coefficient is
\begin{equation}
  \label{eq:deff}
  2D_\text{eff}\equiv\lim_{\TT\to\infty}\TT(\mean{J_\TT^2}-\mean{J_\TT}^2) =  \alpha''(0).
\end{equation}
For better readability, we will refer to the scaled cumulant generating function as the ``generating function'' in the following. 

The generating function
\eqref{eq:defGenFunc} can be calculated as the largest real eigenvalue of a
modified Fokker-Planck operator \eqref{eq:FokkerPlanck}, called the tilted
operator. For the problem at hand the tilted operator
reads 
\begin{equation}
	\label{eq:tiltedOperator} \mathcal{L}(\lambda) = \mathcal{L} + \lambda v .
\end{equation}
Details on the derivation of the tilted operator can be found in the
appendix. 

The right and left eigenfunctions of $\mathcal{L}(\lambda)$ corresponding to the
largest eigenvalue $\alpha(\lambda)$ will be  denoted
as $g(\ps,\lambda)$ and $\tilde g(\ps,\lambda)$, respectively. The
tilted operator satisfies the identity
\begin{equation}
  \label{eq:operator_gc}
  \mathcal{S}\mathcal{L}^\dagger(\lambda)\mathcal{S}=\mathcal{V}^{-1}\mathcal{L}(-\lambda-\fext/T)\mathcal{V}
\end{equation}
with the diagonal operator $\mathcal{V}\equiv \exp[-E(\ps)/T]$
and the operator $\mathcal{S}$ just flipping the sign of $v$. This identity
implies the Gallavotti-Cohen symmetry \cite{kurc98}
\begin{equation}
	\label{eq:galavottiCohen} \alpha(\lambda) = \alpha(-\lambda -{\fext}/{\kT} ),
\end{equation}
which is expressed in terms of the LDF as
\begin{equation}
  \label{eq:FT}
  I(-J)=I(J)+\fext J/T.
\end{equation}
Moreover, Eq.~\eqref{eq:operator_gc} relates the left and right
eigenvectors as
\begin{equation}
  \label{eq:ev_gc}
  \tilde g(\ps,\lambda)=\exp[E(\ps)/T]\,g(x,-v,-\lambda-\fext/T).
\end{equation}

The left and right eigenfunctions can be used to construct the typical densities yielding a specific macroscopic current \cite{laza15}. The product of the eigenfunctions
\begin{equation}
	\label{eq:typicalDensity} \rho_\text{typ}(\ps, \lambda) = \tilde{g}(\ps, \lambda)g(\ps, \lambda)/\mathcal{N}
\end{equation}
with appropriate normalisation $\mathcal{N}$
is the typical density minimizing the functional $I[\rho]$ in equation
\eqref{eq:underdampedLevel20Contraction} for the current
$J=\alpha'(\lambda)$. This distribution is the phase space density associated 
with the ensemble of trajectories producing a specific fluctuation $J_\mathcal{T} = J$. Notably, we have 
$\rho_\text{typ}(\ps,0)=p^\text{s}(\ps)$ and $\rho_\text{typ}(\ps,-\fext/T)=p^\text{s}(x,-v)$.
By solving the eigenvalue problem of the tilted operator \eqref{eq:tiltedOperator} we 
can thus reconstruct the LDF $I(J)$ by means of the G\"artner-Ellis 
theorem \eqref{eq:gaertnerEllis} as well as the typical
phase space density conditioned on the current $J$ corresponding  to $\lambda$. 

In order to numerically calculate the eigenvalue, we discretise the
operator $\mathcal{L}(z)$. Inspired by previous work considering steady state
distributions \cite{risken, lind16}, we expand the
right eigenfunction basis of the tilted operator \eqref{eq:tiltedOperator} in a
Fourier-Hermite basis
\begin{equation}
	\label{eq:fourierHermiteBase} r_{p}^n(\ps) \equiv \frac{\sqrt{\kT}}{2\pi R\sqrt{m}} e^{inx/R}\phi_0(v)\phi_p(v) , 
\end{equation}
consisting of Fourier modes in $x$ and Hermite functions $\phi_p(v)$ in $v$. The Hermite functions are defined as 
\begin{align}
	\phi_p(v)& \equiv \frac{(-1)^p}{\sqrt{2^p p! \sqrt{\pi} }} \left(\frac{m}{\kT}\right)^{\frac{p}{2}} e^{mv^2/(2 \kT)} \frac{\mathrm{d}^p}{\mathrm{d}v^p} e^{-mv^2/\kT} \nonumber \\
	& \equiv H_p(v) \frac{1}{\sqrt{2^p p! \sqrt{\pi}}} e^{-mv^2/(2 \kT)}  
\end{align}
where the second line introduces the Hermite polynomials $H_p(v)$. 
For the left eigenfunction, we use the basis
\begin{equation}
	\label{eq:fourierHermiteBaseLeft} l_{p}^n(\ps) \equiv  \frac{1}{\sqrt{2^p p! \sqrt{\pi}}} e^{inx/R} H_p(v)
\end{equation}
with Hermite polynomials $H_p(v)$ instead of Hermite functions. This has the
advantage that the left eigenfunction $1$ for $\lambda=0$ can be trivially
represented using only the zeroth order Hermite polynomial. The basis vectors
$l_p(\ps)$ and $r_p(\ps)$ are orthogonal and normalised, such that the matrix
elements of the tilted operator are given by the integral
\begin{equation}
	\mathcal{L}(\lambda)_{pp'}^{nn'} = \psInt l_{p}^n(\ps) \mathcal{L}(\lambda) r_{p'}^n (\ps).
\end{equation}
Since the $v$-derivatives and -factors in the tilted operator only cause an
index shift of the right basis vectors, most of the matrix elements vanish. As
a result, $\mathcal{L}(\lambda)$ becomes a tridiagonal block matrix with
entries
\begin{align}
	\label{eq:tiltedOperatorHermiteFourierBasis}
	\bm{\mathcal{L}}(\lambda)_{pp'} = &-\sqrt{p}\left( \uu{\hat{D}} - \lambda \sqrt{\frac{\kT}{m}}  \uu{I} \right) \delta_{p-1,p'} - p\frac{\gamma}{m} \delta_{p,p'}   \nonumber \\
	& -  \sqrt{p+1}\left( \uu{D} - \lambda \sqrt{\frac{\kT}{m}} \uu{I}\right) \delta_{p+1,p'}
\end{align}
with the identity matrix $\uu{I}$ and the matrices
\begin{align}
	\hat{D}_{n n'} &\equiv \left( \delta_{n,n'}\left( in \kT - F_\text{ext} \right)  + \tilde{F}_{nn'} \right) /\sqrt{\kT m} \\
	D_{n n'} &\equiv  in  \delta_{n,n'} \sqrt{\kT/m}
\end{align}
where
\begin{equation}
  \tilde{F}_{nn'}  \equiv  \int_0^{2\pi R} dx \, e^{i(n'-n)x/R} V'(x) 
\end{equation}
 is the Fourier transform matrix of the potential force acting on the particle. For the simple case of a cosine potential
\begin{equation}
	\label{eq:cosinePotential}V(x) = V_0\cos(x),
\end{equation}
the Fourier transformation  yields
\begin{equation}
	\tilde{F}_{nn'}  = \frac{iV_0}{2} \left( \delta_{n,n'-1} - \delta_{n,n'+1}\right) .
\end{equation}
The same expression for the discretized tilted operator \eqref{eq:tiltedOperatorHermiteFourierBasis} has also been derived in the context of Josephson junctions using a slightly different approach in \cite{zond15}.

\section{Numerical case study}
\label{sec:numerics}

\begin{figure}
	\includegraphics{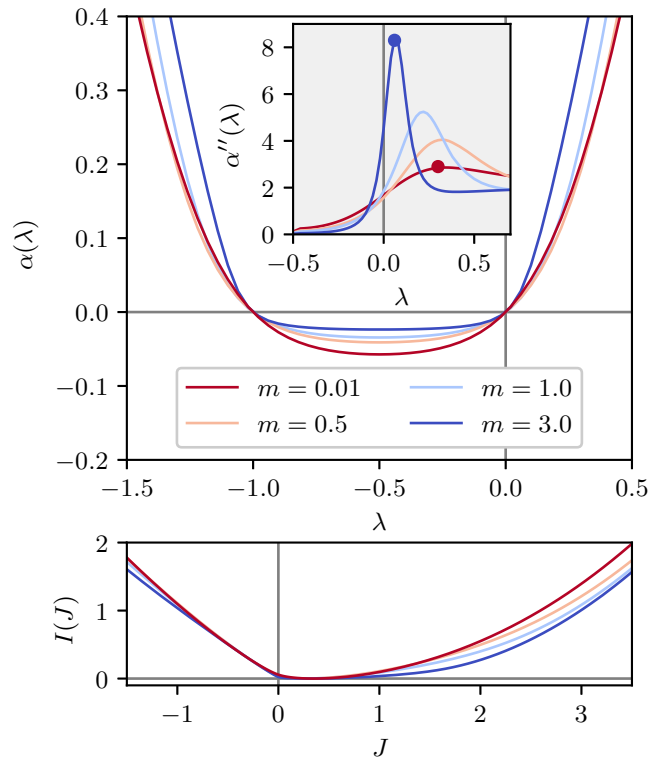}	
	\caption{\label{fig:generatingFunction} Numerical results for the
          underdamped motion in the cosine potential. The upper panel shows the
          generating function $\alpha(\lambda)$ for different masses with
          $T=1$, $\gamma=1$ and $V_0 = 2$, $\fext=1$. The second derivative is
          shown in the inset, with the points of largest curvature marked by
          coloured dots.  The lower panel shows the corresponding LDF $I(J)$
          calculated via the Legendre-Fenchel transform \eqref{eq:gaertnerEllis}. }
\end{figure}

For a numerical case study, we focus on the underdamped motion in the cosine
potential \eqref{eq:cosinePotential}.
The eigenproblem to the transformed operator
\eqref{eq:tiltedOperatorHermiteFourierBasis} can be solved numerically by
truncating after $64$ Hermite functions and limiting the Fourier modes to
$-80 \leq n \leq 80$.
Furthermore, we can employ the Gallavotti-Cohen symmetry
\eqref{eq:galavottiCohen} of the generating
function to restrict ourselves to $\lambda \geq -\fext/2\kT$.
Figure \ref{fig:generatingFunction}
shows the generating function for three different masses with $T=1$ and
$\fext=1$. As
already reported for overdamped motion \cite{tsob16}, the generating function
displays a plateau around the centre of symmetry with little curvature. Away
from this plateau, the generating function converges towards a parabola with
the characteristics of free diffusion. In particular the curvature equals
$\alpha''(\lambda)=2D$ for $|\lambda| \gg 0$ with the bare diffusion
coefficient $D\equiv T/\gamma$. 

\begin{figure*}
		\includegraphics[scale=1]{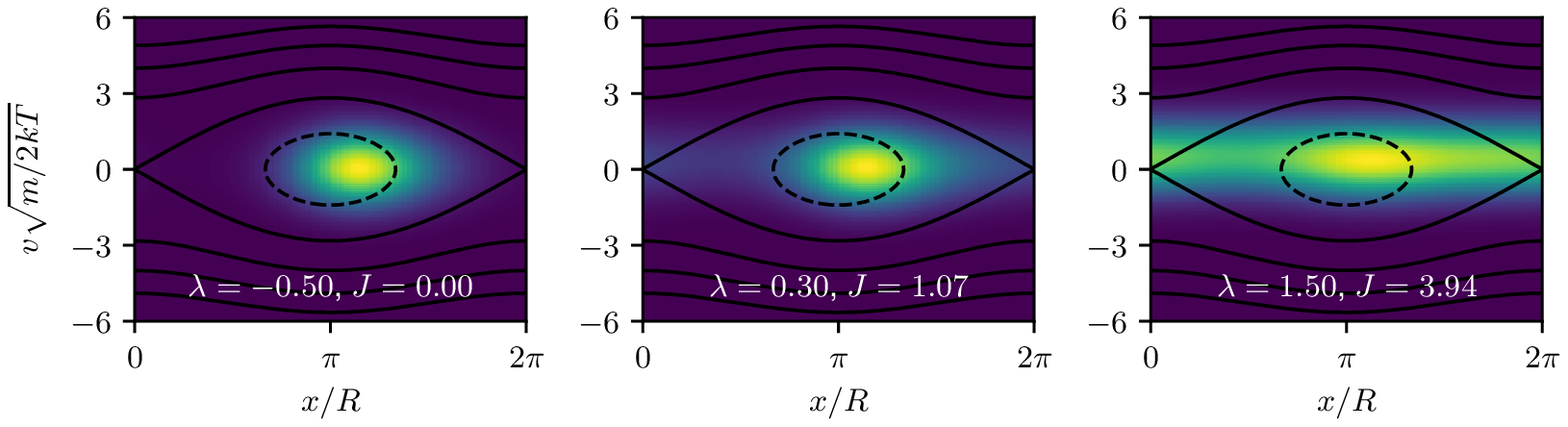}	
		\includegraphics[scale=1]{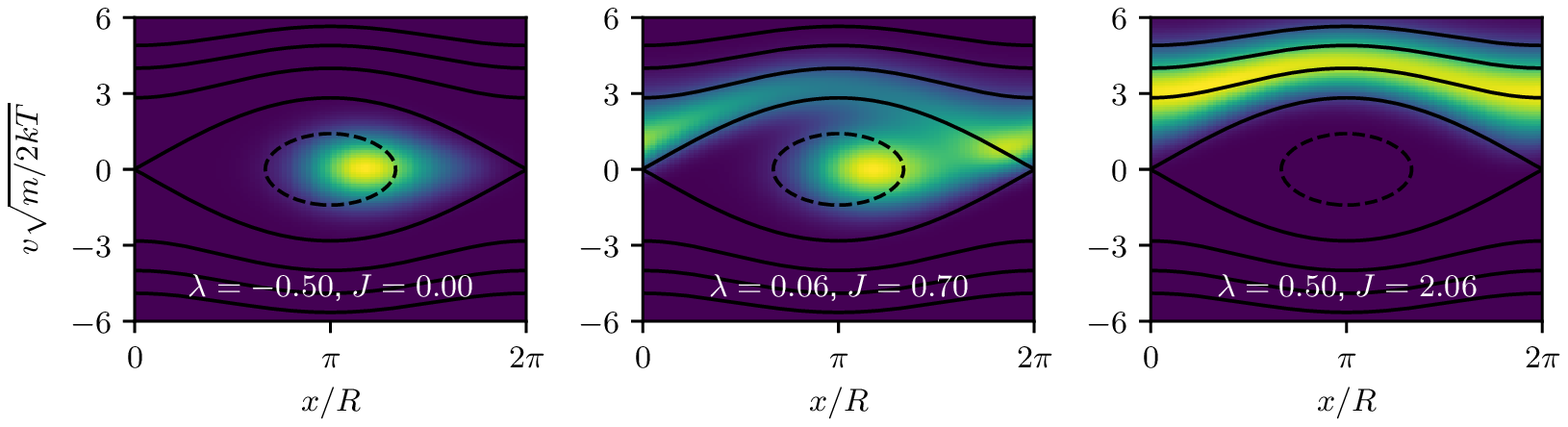}
                \caption{\label{fig:distributions} Typical densities for
                  small mass $m=0.01$ (top row) and large mass $m=3$ (bottom
                  row) with $\kT=1$, $\gamma=1$ and $V_0 = 2$. Darker colour
                  correspond to a small probability density. The tilting
                  $\lambda$  of the operator \eqref{eq:tiltedOperator} and with it the empirical current $J$ is
                  varied along the columns as specified at the bottom of each
                  plot. The contour lines of the energy landscape are plotted
                  as black lines. The first column corresponds to the state with
                  vanishing current $J = 0$. }
\end{figure*}

The two regimes of the generating function can be associated with
different properties of the typical distributions $\rho_\text{typ}(\ps,\lambda)$,
Eq.~\eqref{eq:typicalDensity}, as shown in figure \ref{fig:distributions}. 
For $\lambda=-1/2$, shown in the first column, the typical distribution is
symmetrical in $v$, as evident from Eqs,~\eqref{eq:ev_gc} and \eqref{eq:typicalDensity}. 
The mean particle current associated with this distribution vanishes.
In the vicinity of this value of $\lambda$, the distributions of the velocity
are approximately Gaussian with mean close to zero  for all $x$. Along the  $x$ direction 
the densities show a local maximum at approximately $\pi$. Around $x=0$ the probability
is close to zero. The trajectories producing this phase-space density  are ``locked in'' by the potential 
landscape and rarely cross the potential barrier.

As the tilting parameter $\lambda$ increases to larger values, the particle
flux increases as well, which gives rise to stripes of elevated probability
spreading over the complete $x$ range as shown in the third column. These stripes correspond to
``running'' trajectories that can overcome the potential barrier. Even
though these stripes occur at large tilting for both large and small masses, they
show a mass-dependent characteristic. While for small masses the probability
distribution resembles a Gaussian distribution in $v$ with little dependence
on $x$, the stripes tightly follow the contour lines of the internal energy
$E(\ps)$ for large masses. This behaviour is due to the fact that the noise intensity in $v$ scales
like $1/m$, leading to only small deviations from deterministic trajectories for large mass. 
For small mass, in contrast, the relaxation time becomes small thus inhibiting the 
memory effects of energy conservation. 

In an intermediate regime the typical densities display both characteristics,
a ``running'' stripe and the ``locked'' local maximum around $v=0$, corresponding
to ``running'' and ``locked'' parts of trajectories, respectively.  The value of $\lambda$ in this
intermediate regime coincides with the position $\lambda_\text{c}$ of the maximum in the second
derivative of the generating function, marked with dots in the inset of figure
\ref{fig:generatingFunction}.  The maximum of the second derivative marks the transition 
from the flat plateau to the quadratic regime. The plateau of the generating function is 
consequently bound between $-\lambda_\text{c}-\fext/T$ and $\lambda_\text{c}$. 
With increasing mass, the intermediate regime
becomes narrower, leading to a larger second derivative of the generating function 
around $-\lambda_\text{c}-\fext/T$ and $\lambda_\text{c}$. 

By Legendre-Fenchel transformation, the plateau of the generating function
translates to a kink in the LDF at $J=0$, which becomes sharper when the  second derivative at 
$\lambda_\text{c}$ becomes larger. Such a kink in the LDF is a common feature and
has been observed and discussed in a wide variety of systems such as models describing
driven overdamped systems \cite{mehl08} and molecular motors \cite{laco08}. 
In some cases, the kink in the LDF has been attributed to
intermittent or flashing states \cite{spec12, budi11a}. In the underdamped system, the
bistability between ``running'' and ``locked'' dynamics causes such an
intermittence. The increased effective diffusion for intermittent trajectories, which equals
the second derivative of the generating function at $\lambda = 0$, 
suggests that the phenomenon of enhanced diffusion \cite{reim01} is observed
when locked and running trajectories have a comparable influence on the motion.

\begin{figure*}
	\includegraphics{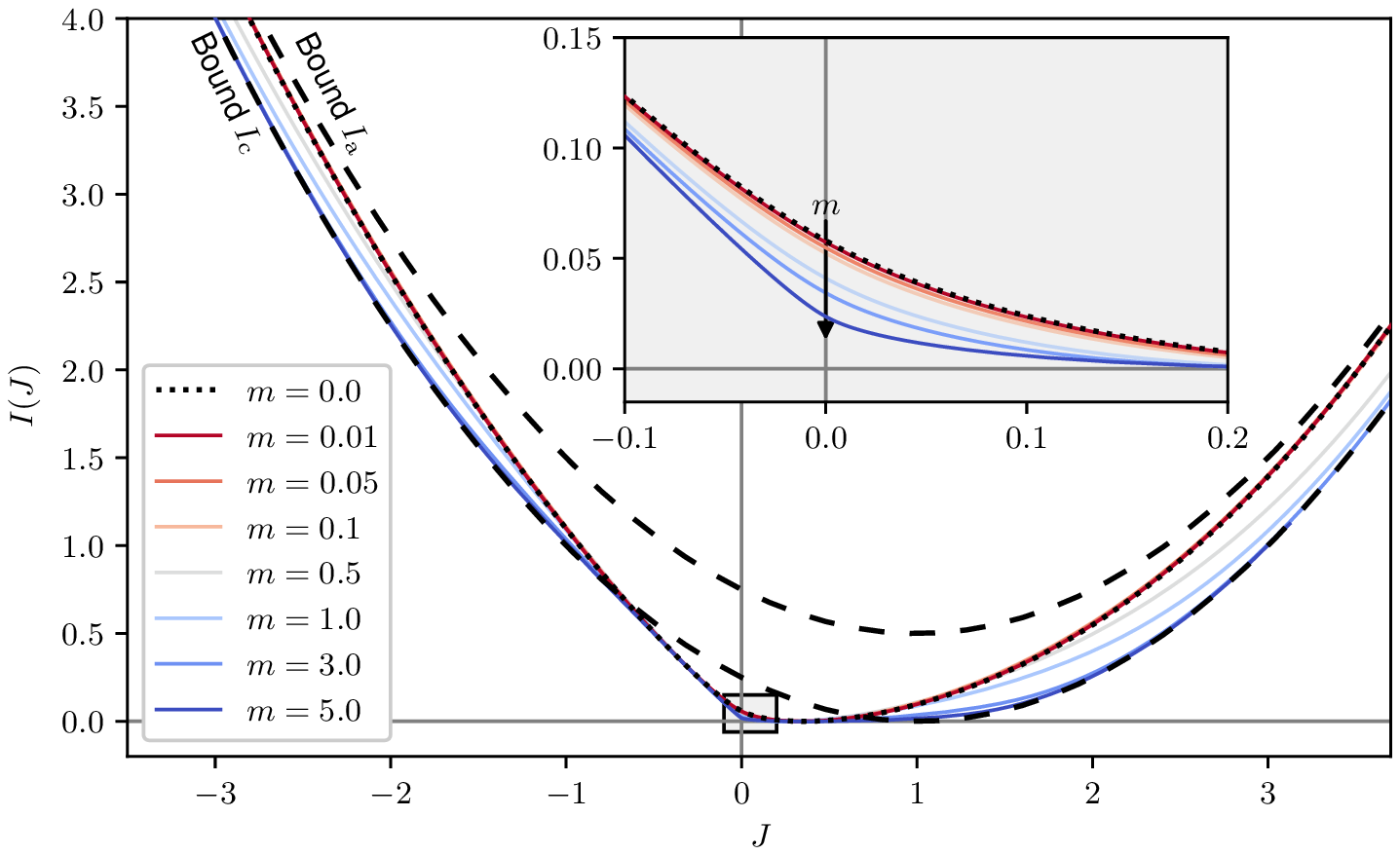}
	\caption{\label{fig:numericalResults} LDF for the cosine potential for
          different masses $m$ showing the transition from underdamped to
          overdamped motion. The inset displays the functions in the vicinity
          of $J=0$ as indicated by the grey box. The dashed lines correspond
          to the asymptotic bounds $I_\text{a}$ and $I_\text{c}$, where the former one
          is valid for the overdamped case $m=0$ and the latter for the extreme 
          underdamped case
          $m \rightarrow \infty$. The other parameters are $V_0=2.0$, $\kT=1$,
          $\gamma=1$, $\fext=1$.}
\end{figure*}

The sharpening of the kink for increasing mass at $J=0$,
as displayed in Fig.~\ref{fig:numericalResults}, can be interpreted as a
tightening regime of intermittent states. This is due to the fact that a
particle with larger inertia survives longer in a running state compared to a
lighter particle. Hence, the intermittent switching between running and trapped
trajectories becomes more sensitive for the tilting $\lambda$ for large masses.

Furthermore, the LDF seems to converge to limiting functions for both large
and for small masses. For small masses, the exact overdamped LDF is reproduced. 
For large masses, the LDF becomes quadratic for large empirical currents with a 
pronounced plateau between $0$ and the stationary current $J^\text{s}$.

\section{Bounds on the LDF}
\label{sec:bounds}

Bounds on the LDF can be constructed from the contraction
\eqref{eq:underdampedLevel20Contraction} by inserting trial densities
tailored to produce a specific particle current. The constraint of the
empirical current \eqref{eq:underdampedEmpiricalCurrent} in the underdamped
functional LDF, however, poses a new problem for finding bounds on the LDF
for the particle current, since it does not allow for a variation of the
current independently from the density, as was possible in the 
overdamped case \cite{ging16,pole16}. Instead, a
suitable ansatz has to be developed for the complete phase space density. In
the following, we provide some trial functions, which prove useful for the
determination of upper bounds on the LDF.

\subsection{A master ansatz}
A versatile ansatz for the empirical density is the function
\begin{align}
	\label{eq:versatileAnsatz} \rho(\ps; J) &= \rho(x ) \sqrt{\frac{m}{2\pi \kT}} \exp\left[ -\frac{m}{2\kT} \left( v - \frac{J}{2\pi R \rho(x )}\right)^2 \right] \nonumber \\
	& \equiv  \rho(x ) \hat{\rho}(\ps)
\end{align}
with an arbitrary normalized distribution $\rho(x)$. This choice corresponds to a
Gaussian distribution for fixed $x$. The variance of the distribution matches
the equilibrium distribution of the velocity, the centre is at $J/(2\pi R
\rho(x))$. By construction, the macroscopic current \eqref{eq:intcurr} is
$J$. The advantage of the ansatz is that the associated empirical current
$\mu_v^\rho(\ps)$, Eq.~\eqref{eq:underdampedEmpiricalCurrent}, can be cast in
the closed form
\begin{equation}
	\label{eq:versatileCurrent} \mu^\rho_v (\ps) = \rho'(x)\hat{\rho}(\ps) \left[ \frac{\kT}{m} - \frac{Jv}{2 \pi R \rho(x)}\right] .
\end{equation}
Plugging these terms in the functional LDF \eqref{eq:underdampedLevel20LDF}
and performing the contraction with respect to $\rho(x)$
yields, after some simplifications, the upper bound
\begin{widetext}
\begin{align}	
	\label{eq:versatileBound} I(J) \leq I_\text{M}(J) \equiv &\min_{\rho(x)} \frac{1}{4 \gamma \kT} \int dx\, \left\lbrace  \frac{1}{\rho(x)}\left[ \kT \rho'(x) - F(x) \rho(x) + \gamma \frac{J}{2\pi R} \right]^2    \right. \\
											& \left. + m \left( \frac{J}{2\pi R} \right) ^2 \frac{\rho'(x)}{\rho(x)^3 } \left[ 2 \rho(x) F(x) - \kT \rho'(x) \right] +  m^2 \left( \frac{J}{2\pi R} \right) ^4 \frac{\rho'(x)^2}{\rho(x)^5}  \right\rbrace  \nonumber 
\end{align}
\end{widetext}
with the total force $F(x)\equiv\fext-V'(x)$.
In the following, we discuss some special cases and implications of this bound.

\subsection{Overdamped asymptotic bound}
The simplest realization of the ansatz \eqref{eq:versatileAnsatz} is to assume a
uniform distribution $\rho(x) = 1/2\pi R$. This is motivated by the typical
distributions shown in the top row of Fig.~\eqref{fig:distributions}. For large
tilting at small mass, the stripe of elevated probability becomes independent of $x$
with a Gaussian, equilibrium-like distribution in $v$.

Since $\rho(x)$ is flat, the first derivative vanishes as well as the associated
current \eqref{eq:versatileCurrent}. Consequently, most of the terms in
$I_\text{M}(J)$ vanish. For the cosine potential
\eqref{eq:cosinePotential}, this bound $I_\text{a}$ reads 
\begin{equation}
	\label{eq:simpleBound} I(J)\leq I_\text{M} (J) \leq I_\text{a}(J) \equiv
        \frac{\gamma}{4 \kT }\left( J - \frac{\fext}{\gamma}\right) ^2 +
        \frac{V_0^2}{8\kT\gamma} .
\end{equation}
The bound is plotted in Fig.~\ref{fig:numericalResults}. As expected, it
becomes tight for $m \rightarrow 0$ in the tails of the LDF. For larger
masses, however, there is an offset between the bound and the LDFs. This
discrepancy indicates that the assumptions made for this simple ansatz are a
good description for strongly directed overdamped motion, but not for larger
masses. This is due to the fact that the velocity equilibrates for strong
damping on a time scale that is smaller than the time scale for the motion in
the potential. Nevertheless, the bound holds for arbitrary masses and can thus
be regarded as a generalisation of the asymptotic bound reported in
\cite{tsob16} for the overdamped limit.

\subsection{LDF in the overdamped limit}
The underdamped LDF converges to the overdamped LDF in the limit of small masses, as can
be observed in Fig.~\ref{fig:numericalResults}. In this subsection, we examine
how the typical distributions behave in this limit, where the two-dimensional
empirical densities reduce to effectively one-dimensional ones.

The  overdamped LDF of a particle on a ring can be calculated via the contraction \citep{hopp16}
\begin{equation}
	\label{eq:overdampedContraction} I_\text{O}(J) = \min_{\rho(x)} \frac{1}{4 \kT \gamma}  \int dx\, \frac{\left[ (\kT \partial_x  - F(x)) \rho(x) + \frac{\gamma J}{2\pi R} \right]^2}{\rho(x)}  .
\end{equation}
This overdamped contraction can be identified as the limit $m \rightarrow 0$
in $I_\text{M}(J)$ in Eq.~\eqref{eq:versatileBound}. Consequently, the density
ansatz \eqref{eq:versatileAnsatz} with the overdamped empirical distribution $\rho(x)$,
the optimal distribution from the contraction
\eqref{eq:overdampedContraction}, converges to the exact phase space density
in the limit $m\rightarrow 0$.

The tightness of our versatile ansatz for overdamped motion can be justified by
interpreting the ansatz itself. The local mean velocity in the overdamped
limit is given by $J/(2\pi R \rho(x))$. The Gaussian distribution of the velocity reflects the local
equilibrium, for which we adapt the mean to this value.

For the special case $J=0$, the bound $I_\text{a}(J)$, Eq.~\eqref{eq:versatileBound}, becomes
independent of the mass and thus equal to the overdamped limit $I_\text{O}(J)$, i.e.
\begin{equation}
	I(0) \leq  I_\text{O}(0)  = \lim_{m\rightarrow 0} I(0) .
\end{equation}
The probability of the fluctuation $J=0$ is therefore for an underdamped
particle larger than for an overdamped particle with the same friction 
coefficient and temperature.  This behaviour is plotted in the inset of figure
\eqref{fig:numericalResults}. Furthermore, we observe numerically that the
decrease of the LDF $I(0)$ with increasing mass is monotonic.

\subsection{Underdamped asymptotic bound}
Although designed as an asymptotic bound, $I_\text{a}(J)$ in Eq.~\eqref{eq:simpleBound}
is not saturated by the LDF in the tails for large mass, as shown in
Fig.~\ref{fig:numericalResults}. In this limit, the impact of thermal noise on
the velocity is marginal. As a result, we observe approximate energy
conservation, leading to trajectories that mainly follow the contour lines of
the internal energy $E(\ps)$, as depicted in figure \ref{fig:distributions}. The
marginalised probability in $x$, i.e. the mean time spent at a certain
position $x$, is proportional to the inverse mean velocity at this
position. Following this rationale we make the ansatz
\begin{equation}
	\label{eq:underdampedAnsatz} \rho(x, v; E) = \frac{\sqrt{m/(2\pi
            T)}}{N(E) \bar{v}(x;E)} \exp\left[ -\frac{m}{2\kT}\left(  v - \bar{v}(x; E) \right)^2 \right],
\end{equation}
where $\bar{v}(x;E)$  is the local mean velocity and follows the contour line of
the internal energy with level $E$
\begin{equation}
	\bar{v}(x; E) = \sqrt{ \frac{2}{m}\left(E- V(x) + V(0)\right)}, 
\end{equation}
and $N$ is the normalisation
\begin{equation}
	N(E) = \int dx\, \bar{v}(x; E)^{-1} .
\end{equation}
The ansatz~\eqref{eq:underdampedAnsatz} reduces to \eqref{eq:versatileAnsatz} for the choice 
\begin{equation}
	\rho (x; E) = \frac{1}{N(E) \bar{v}(x; E)}\;\; , \;\; J(E) = \frac{2\pi R}{N(E)}.
\end{equation}
Using these terms as trial function for the minimisation in Eq.~\eqref{eq:versatileBound} gives the bound
\begin{align}
	\label{eq:underdampedBound} I(J) \leq& I_\text{M}(J(E))\leq I_\text{b}(J(E)) \nonumber \\
	\equiv & \frac{1}{4\kT \gamma N(E)} \int dx \, \left\lbrace \frac{1}{\bar{v}(x; E)}\left( F - \gamma \bar{v}(x; E ) \right)^2  \right.  \nonumber \\
	& \left. + \frac{\kT}{m}\frac{V'(x)^2}{\bar{v}(x; E)^3} \left(  \frac{\kT}{m \bar{v}(x; E)^2}  + 1\right) \right\rbrace .
\end{align}
This bound can be evaluated for arbitrary masses using numerical integration
schemes. In the limit $m \rightarrow \infty$ with $J(E)$ kept fixed, the energy
$E$ scales linearly in $m$ and $\bar v(x;E)$ becomes independent of $x$. 
The bound then reduces to the simpler expression
\begin{equation}
	\label{eq:underdampedBoundLimit} \lim_{m \rightarrow \infty} I\left( J
        \right) \leq \lim_{m \rightarrow \infty} I_\text{b}\left( J \right)
        = \frac{\gamma}{4\kT} \left( J - \frac{\fext}{\gamma}\right)^2 \equiv I_\text{c}(J), 
\end{equation}
which is the same quadratic function as for the overdamped asymptotic bound
\eqref{eq:simpleBound} up to the minimum being shifted to $0$. 

The quality of the underdamped asymptotic bound as an approximation of the
actual LDF in the limit of large mass can be appreciated in figure
\ref{fig:numericalResults}. For an intermediate mass of $m=3$, the bound
$I_\text{c}(J)$ already matches the tails of the LDF. In this regime, the ansatz
\eqref{eq:underdampedAnsatz} describes well the typical distributions
resulting from running trajectories with high energy that have only small
modulations in the velocity.

\section{An underdamped thermodynamic uncertainty relation?}
\label{sec:tdur}

In order to characterize the typical particle currents, it is essential to
find differentiable bounds on the LDF that are saturated in the vicinity of
its minimum at $J=J^\text{s}$.  Such a bound implies locally a bound on the
effective diffusion coefficient in Eq.~\eqref{eq:deff}, which is also given by
$2D_\text{eff}=1/I''(J^\text{s})$. A prominent example is the quadratic bound \cite{piet15}
\begin{equation}
  \label{eq:quadbound}
  I(J)\leq \frac{\sigma}{4}(J/J^\text{s}-1)^2,
\end{equation}
which involves the entropy production rate $\sigma$ and implies the
thermodynamic uncertainty relation \cite{bara15} and holds (at least) for
overdamped Brownian dynamics and continuous time Markovian jump processes. The
proof under these premises relies on the contraction principle for the level
2.5 LDF, where one is free to chose an ansatz for the empirical density and,
independently, for the empirical current \cite{ging16,pole16,ging16a}. In that proof, 
the empirical density was chosen equal to the stationary density, while the 
empirical current was scaled independently. This choice can be interpreted as 
a ``time-lapse'' transformation of steady state trajectories, where all trajectories are
scaled linearly in time. Such an ensemble of
``time-lapsed'' trajectories reproduces the stationary density but leads to 
a scaled empirical current.

Following the same idea for underdamped Brownian dynamics and rescaling the
trajectories by the factor $(1+c)$ leads to the ansatz
\begin{equation}
	\label{eq:timelapseAnsatz} \rho(\ps; c) =  p^\text{s}(x,  v/(1+c))/(1+c)
\end{equation}
for the phase-space density. By construction, the marginalized $x$-distribution
of this density is the same as for the stationary distribution and the 
particle current~\eqref{eq:level2current} is scaled as
\begin{equation}
	\label{eq:timelapseCurrent} J = \psInt v \rho(\ps; c) = (1+c) J^\text{s} .
\end{equation}
Plugging the trial function into the contraction
\eqref{eq:underdampedLevel20LDF} and expanding around the stationary state at $c = 0$ gives the bound
\begin{align}
	\label{eq:timelapseBound}  I\left( J \right) \leq
        I_\text{TL}\left(J\right)&\equiv I[ \rho(\ps; c=J/J^\text{s}-1)] \\
	 & = Ac^2  + \mathcal{O}(c^4), \nonumber
\end{align}
with
\begin{equation}
  \label{eq:DTL}
  A\equiv\left[ \frac{5}{4}\sigma + \frac{3\gamma}{m} - \frac{2\gamma}{\kT} \langle v^2 \rangle + \frac{1}{\kT \gamma} \langle F(x)^2 \rangle \right],
\end{equation}
implying $D_\text{eff}\geq (J^\text{s})^2/(4A)$ as bound on the effective diffusion coefficient.

In contrast to the quadratic bound for overdamped motion \eqref{eq:quadbound},
the bound \eqref{eq:timelapseBound} has typically a much larger curvature at
$J=J^\text{s}$ and is saturated neither in the limit of a large external force nor
in the linear response regime. 

Under which conditions the bound \eqref{eq:quadbound} holds for underdamped dynamics thus 
remains an open question. In the numerical case studies presented here, we have
not seen any violation of \eqref{eq:quadbound}, however, the high
computational cost to calculate large deviation functions for arbitrary
potentials and geometries prohibits extensive numerical checks.

It seems likely that the bound \eqref{eq:quadbound} can be violated if a
magnetic field breaks the micro-reversibility of the trajectories, as has been
observed for piecewise deterministic particle transport between
reservoirs \cite{bran18}.

In Ref.~\cite{piet17}, it has been shown that in the linear response regime,
the bound \eqref{eq:quadbound} follows as a consequence of a fluctuation
relation. Since this fluctuation relation builds only on micro-reversibility,
one can conclude that \eqref{eq:quadbound} holds for underdamped dynamics in
the linear response regime and in the absence of magnetic fields.

A full proof of the bound \eqref{eq:quadbound} for underdamped dynamics would
probably require large deviation techniques that go beyond the established
methods. Insight might also come from martingale methods \cite{pigo17}, which
at the current stage still requires the diffusion tensor to be invertible,
which is not the case for underdamped diffusion in phase space.

\section{Conclusion}
\label{sec:conclusion}

\begin{figure}
	\includegraphics[scale=1]{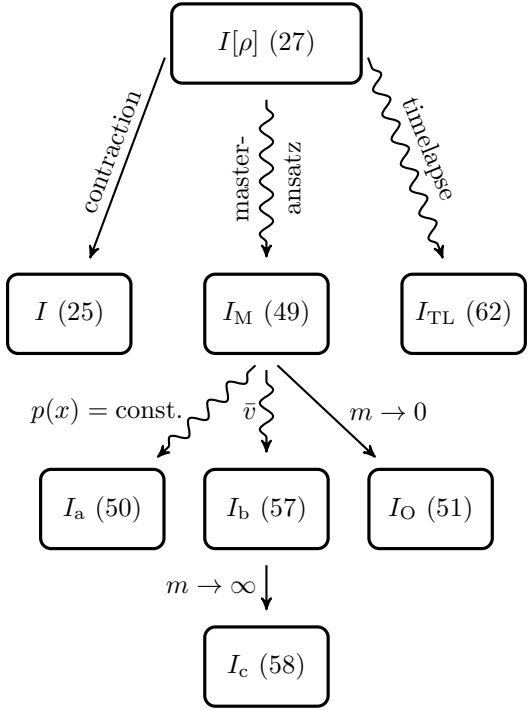}
\caption{Hierarchy of the large deviation functions and bounds discussed in
  this paper. Straight arrows denote equalities due to full contraction or
  specification for limiting cases; wavy arrows point to upper bounds due to
  non-optimal trial functions for the contraction.}
\label{fig:tree}
\end{figure}

For the driven underdamped Brownian motion of a particle in one dimension, we have
derived in Sec.~\ref{sec:level2} the level 2 large deviation function, which
characterises the probability to observe an empirical phase space
density. This result replaces the level 2.5 large deviation function for the
joint probability of densities and currents, that is typically used for
overdamped Brownian motion and Markov jump processes. The large deviation function characterising
the fluctuations of the particle current can be either calculated through
contraction from this level 2 large deviation function or, as described in
Sec.~\ref{sec:scgf}, via Legendre transformation of the cumulant generating
function. While the latter method is suitable for a numerical computation of the
large deviation function in Sec.~\ref{sec:numerics}, the former method allows one
to derive bounds on the large deviation function by inserting tailored trial
functions for the typical phase space densities. Fig.~\ref{fig:tree} summarizes the 
hierarchy of bounds that we have obtained following this principle. 

In Sec.~\ref{sec:bounds}, we have derived several bounds from a ``master
ansatz'' that eliminates the velocity in the contraction principles. This
ansatz reproduces the large deviation function for overdamped dynamics $I_\text{O}(J)$ exactly.
Additionally choosing a constant distribution in the position yields $I_\text{a}(J)$
as a simple bound, which becomes asymptotically tight for small inertia and
large fluctuations. Another bound $I_\text{b}(J)$, employing a mean velocity adapted
to the position, becomes particularly tight in the limit of large inertia,
where it assumes a quadratic form $I_\text{c}(J)$.

A different bound $I_\text{TL}(J)$ has been derived in Sec.~\ref{sec:tdur}
based on a ``time-lapse'' transformation of the stationary phase space
density. While this procedure proves the thermodynamic uncertainty for
overdamped dynamics and Markov jump processes, we have shown that
$I_\text{TL}(J)$ implies for underdamped dynamics a different, typically
weaker bound on the effective diffusion coefficient.

We have focused on the underdamped dynamics of an isothermal, one dimensional
system with time-independent driving. This has the advantage that the particle
current, being proportional to the dissipated heat, is the only relevant
macroscopic current. It will be interesting to generalise our approach to more
complex situations, such as models for heat engines with different input and
output currents \cite{blan98,verl14,piet17a}. A generalisation of our results
to fluctuations on finite time-scales \cite{horo17} and first passage time
fluctuations \cite{ging17} should be feasible.

\appendix

\section{\label{sec:appDerivTiltedOp} Derivation of the tilted operator}

In the following we derive the tilted operator \eqref{eq:tiltedOperator} in
full phase space, Eq.~\eqref{eq:tiltedOperator}, following the standard
methods as pedagogically reviewed in \cite{touc17}. To this end we introduce
an additional variable $y$ which corresponds to the integrated, travelled
distance $\TT J_\TT$. The variable $y$ is not limited in the interval
$[0, 2\pi R)$, in contrast to the variable $x$ losing information on the total travelled distance. We are
interested in the generating function
\begin{equation}
	\label{eq:genFunc} g(x, v, \lambda, t) \equiv \langle e^{\lambda y}
        \rangle = \int_{-\infty}^{\infty} e^{\lambda y} p(x, v, y, t)\, dy,
\end{equation}
where the probability function $p(x,v,y,t)$ satisfies the Fokker-Planck equation
\begin{equation}
	\label{eq:genFuncFPE} \partial_t p(x, v, y, t) = \left[\mathcal{L}-\partial_y v\right] p(x, v, y, t)
\end{equation}
with continuous periodic boundary condition in $x$.
This is the Fokker-Planck equation of the original
problem~\eqref{eq:FokkerPlanck} modified by the term $\partial_y v$
accounting for the variable $y$ with the same drift term as $x$. 

The tilted operator $\mathcal{L}(\lambda)$ describes the time evolution of $g(x, y, \lambda, t)$ for
fixed tilting $\lambda$. By plugging in Eq.~\eqref{eq:genFuncFPE} and
the unmodified operator $\mathcal{L}$ (Eq.~\eqref{eq:FokkerPlanck}), one
gets through integration by parts
\begin{align}
	\label{eq:tiltedFPE}\partial_t g(x, y, \lambda, t) = \left( \mathcal{L} + \lambda v\right) g(x, v, \lambda, t) \equiv \mathcal{L}(\lambda) g(x, v, \lambda, t) .
\end{align}

Let $\alpha_n(\lambda)$ denote the eigenvalues of $\mathcal{L}(\lambda)$
sorted by their real part, where the subscript $0$ indicates the eigenvalue
with largest real part. The Perron-Frobenius theorem guarantees that this
eigenvalue is real. Then the generating function in \eqref{eq:tiltedFPE} can be
expanded in the associated eigenfunctions $q_n(x, y, \lambda)$ as
\begin{equation}
  g(x, v, \lambda,t )=\sum_n c_n(t)\,q_n(x, v, \lambda )
\end{equation}
with expansion coefficients determined by
\begin{equation}
	\label{eq:tiltedFPEExpanded} 
        \sum_n c'_n(t) q_n(x, v, \lambda ) =
        \sum_n \alpha_n(\lambda ) c_n(t) q_n(x, v, \lambda ) 
\end{equation}
as
\begin{equation}
  \label{eq:tiltedFPEExpanded2}
	c_n(t) = c_n(0)e^{\alpha_n(\lambda)t}.
\end{equation}
By applying the logarithm and taking the long time limit, one obtains the scaled cumulant generating
function \eqref{eq:defGenFunc} $\alpha(\lambda)=\alpha_0(\lambda)$ as the largest eigenvalue of the tilted operator \eqref{eq:tiltedFPE}.




\input{pre_manuscript.bbl}

\end{document}

%% file: pre_manuscript.bbl
%